\documentclass{ieeeaccess}
\usepackage{cite}
\usepackage{amsmath,amssymb,amsfonts}
\usepackage{algorithm, algorithmic}
\usepackage{graphicx}
\usepackage{textcomp}
 
\usepackage{placeins}
\usepackage{caption,setspace}
\DeclareMathOperator*{\popcount}{popcount}
\DeclareMathOperator{\sign}{sign}
\DeclareMathOperator{\argmax}{argmax}
\DeclareMathOperator{\XNOR}{XNOR}
\DeclareMathOperator{\BatchNorm}{BatchNorm}
\DeclareMathOperator{\softmax}{softmax}
\DeclareMathOperator{\BackBatchNorm}{BackBatchNorm}
\DeclareMathOperator{\Clip}{Clip}
\DeclareMathOperator{\UpdateAdam}{UpdateAdam}
\DeclareMathOperator{\MovingAverage}{MovingAverage}

\def\BibTeX{{\rm B\kern-.05em{\sc i\kern-.025em b}\kern-.08em
    T\kern-.1667em\lower.7ex\hbox{E}\kern-.125emX}}
\begin{document}
\history{Date of publication 2019.}
\doi{TBA}

\title{Stochastic Computing for Hardware Implementation of Binarized Neural Networks }
\author{\uppercase{Tifenn Hirtzlin}\authorrefmark{1},
\IEEEmembership{Student, IEEE},
\uppercase{Bogdan Penkovsky  \authorrefmark{1}}, 
\uppercase{Marc Bocquet \authorrefmark{2}}, 
\uppercase{Jacques-Olivier Klein \authorrefmark{1}}
\IEEEmembership{Member, IEEE}, 
\uppercase{Jean-Michel Portal \authorrefmark{2}} \\
and \uppercase{Damien Querlioz}\authorrefmark{1}
\IEEEmembership{Member, IEEE}}
\address[1]{Centre de Nanosciences et de Nanotechnologies, Univ. Paris-Sud, CNRS, France}
\address[2]{Institut Mat\'eriaux Micro\'electronique Nanosciences de Provence, Univ. Aix-Marseille et Toulon, CNRS, France}
\tfootnote{This work was supported by the European Research Council Starting Grant NANOINFER (715872) and Agence Nationale de la Recherche grant NEURONIC (ANR-18-CE24-0009).
\copyright
2019 IEEE.  Personal use of this material is permitted.  Permission from IEEE must be obtained for all other uses, in any current or future media, including reprinting/republishing this material for advertising or promotional purposes, creating new collective works, for resale or redistribution to servers or lists, or reuse of any copyrighted component of this work in other works.}

\markboth
{Author \headeretal: Preparation of Papers for IEEE TRANSACTIONS and JOURNALS}
{Author \headeretal: Preparation of Papers for IEEE TRANSACTIONS and JOURNALS}

\corresp{Corresponding author: Tifenn Hirtzlin (email: tifenn.hirtzlin@c2n.upsaclay.fr), Damien Querlioz (email: damien.querlioz@c2n.upsaclay.fr)}

\begin{abstract}

Binarized Neural Networks, a recently discovered class of neural networks with minimal memory requirements and no reliance on multiplication, are a fantastic opportunity for the realization of compact and energy efficient inference hardware.
However, such neural networks are generally not entirely binarized: their first layer remains with fixed point input. 
In this work, we propose a stochastic computing version of Binarized Neural Networks, where the input is also binarized. Simulations {on the example of the Fashion-MNIST and CIFAR-10} datasets show that such networks can approach the performance of conventional Binarized Neural Networks. We evidence that the training procedure should be adapted for use with stochastic computing.
Finally,  the ASIC implementation of our scheme is investigated, in a system that closely associates logic and memory, implemented by Spin Torque Magnetoresistive Random Access Memory. This analysis shows that the stochastic computing approach can allow considerable savings with regards to conventional Binarized Neural networks in terms of area ({$62\%$} area reduction on the Fashion-MNIST task). It can also allow important savings in terms of energy consumption, if we accept reasonable reduction of accuracy: for example  a factor {$2.1$} can be saved, with the cost of {$1.4\%$} in Fashion-MNIST test accuracy.
These results  highlight the high potential of Binarized Neural Networks for hardware implementation, and that adapting them to hardware constrains can provide important benefits. 
%of neural networks.
\end{abstract}


\titlepgskip=-15pt

\maketitle

\section{Introduction}
\label{sec:introduction}

\PARstart{R}{ecent} advances in deep learning have transformed the field of machine learning, with numerous achievements  in image or speech recognition, machine translation and others. 
However, a considerable challenge of deep neural network remains their energy consumption, which limits their use within embedded systems  \cite{editorial_big_2018}.
The hardware implementation of deep neural networks is a widely investigated approach to increase their energy efficiency. 
A particularly exciting opportunity is to rely on in-memory or near-memory computing implementations \cite{yu2018neuro,ielmini2018memory,querlioz2015bioinspired,burr2017neuromorphic,giacomin2018robust}, which are highly energy efficient as they avoid the von Neumann bottleneck entirely.
This idea takes special meaning today, in particular with the emergence of novel memories such Resistive and Magnetoresistive Random Access Memories (RRAMs and MRAMs). 
Such memories are fast and compact non volatile memories, 
which can be embedded at the core of CMOS processes,
and therefore provide an ideal technology for realizing  in-memory neural networks \cite{yu2018neuro,ielmini2018memory,burr2017neuromorphic}.

A  considerable challenge of this approach is that modern neural networks require important amounts of memory \cite{canziani2016analysis}, 
which is not necessarily compatible with hardware in-memory computing approaches.
{Multiple roads have been explored to reduce the precision and memory requirements of neural networks. 
The quantization of the weights used for inference is the most natural route \cite{hubara2017quantized}. Architectural optimization can result in considerable reduction in terms of number of parameters and arithmetic operations, with only modest reduction in  accuracy \cite{sandler2018mobilenetv2}. Network pruning  \cite{reagen2016minerva} or network  compression \cite{chen2015compressing,han2015deep} techniques, sometimes combining different methods, can allow implementing hardware neural networks with reduced memory access and therefore higher energy efficiency.}

{
Binarized Neural Networks (BNNs) have recently appeared as one of the most extreme vision of low precision neural networks, as they go further than these approaches
\cite{courbariaux2016binarized,rastegari2016xnor}.}
In these simple deep neural networks, synaptic weights as well as neuron activations assume Boolean values. 
These models can nevertheless achieve state-of-the-art performance on image recognition, while being multiplier-less, and relying only on simple binary logic functions. 
First hardware implementations have already been investigated and have shown highly promising results \cite{bocquet2018memory,nurvitadhi2016accelerating,yu2018neuro,giacomin2018robust}.

However, BNNs are not entirely binarized: the first layer input is usually coded as a fixed point real number. 
This fact is not a significant issue for operating BNNs on graphical processor units (GPUs) \cite{courbariaux2016binarized}, as they feature extensive arithmetic units. 
Research  aimed at implementing binarized neural network on 
Field Programmable Gate Arrays (FPGAs) \cite{zhao2017accelerating} has also not specifically investigated  the question of the non-binarized first layer:  these works usually use the Digital Signal Processors (DSPs) of the FPGA to process the associated operations. 
However, in an application-specific integrated circuits (ASIC) implementation, the non-binarization of the first layer 
{implies that} 
this layer needs a specific design, which is  more energy  consuming 
and uses more area
than the design used for the purely binary layers.

For this reason,  in this work, we introduce a stochastic  computing implementation of BNNs, which allows implementing them in an entirely binarized fashion. The network functions by presenting several stochastically binarized versions of the images to the BNN, in a way reminiscent to the historic concept of stochastic computing \cite{gaines1969stochastic}.
{
After presenting the background of the work (section~\ref{sec:background}), the paper describes the  following contributions.}

\begin{itemize}
\item {We show that this stochastic  computing implementation of  BNNs allows achieving high network performance in terms of recognition rate on the Fashion-MNIST and CIFAR-10 datasets. Stochastic BNN quickly approaches standard BNN performance when several stochastic binarized images are presented to the network.  We also evidence that strategy for training stochastic computing BNNs should differ from the one used for conventional BNNs (section~\ref{sec:network}).}
\item {We design a full hardware ASIC  in-memory BNN, which allows showing that the stochastic   computing BNN strategy can save important area ($62\%$ on Fashion-MNIST) and energy (factor $2.1$ on Fashion-MNIST with an accuract reduction of $1.4\%$  with regards to a standard BNN (section~\ref{sec:hardware}). These numbers are discussed with regards to different alternative implementations.}
\end{itemize}

%%%%%%%%%%%%%%%%%%%%%%%%%%%%%%%%%%%%%%%%%%%%%%%%%%%%%%%%%%%%%%%%%
%%%%%%%%%%%%%%%%%%%%%%%%%%%%%%%%%%%%%%%%%%%%%%%%%%%%%%%%%%%%%%%%%

\section{Background of the Work}
\label{sec:background}

\subsection{Binarized Neural Networks}

In this section, we first introduce the general principles of Binarized Neural Networks, an approach to considerably reduce the computation cost of  inference in neural networks \cite{courbariaux2016binarized,rastegari2016xnor}. 
In a conventional neural network with $L$ layers, the activation values of the neurons of layer $k$, $a^{[k]}_i$, are obtained by applying a non-linear activation function $f$ to the  matrix product between real-valued synaptic weight matrix $W^{[k]}$ and the real-valued activations of the previous layer of neurons $a^{[k-1]}$: 

\begin{equation}
a^{[k]}_i = f \left( \sum_{j}{W^{[k]}_{ij} \cdot a^{[k-1]}_j} \right).    
\label{eq:activation}
\end{equation}

{In a BNN,    
excluding the first layer,} 
neuron activation values  as well as synaptic weights  assume binary values,  meaning $+1$ and $-1$. 
The products between  weights and neuron activation values in Eq.~(\ref{eq:activation})  then simply become { logic XNOR operation.} 
The sum in Eq.~(\ref{eq:activation})  is  replaced by the $\popcount$ operation, the basic function that counts the number of ones in a data vector. 
The resulting value is then converted to a binary value by comparing it to a trained threshold value $\mu^{[k]}_i$. Eq.~(\ref{eq:activation}) therefore becomes:

\begin{equation}
a^{[k]}_i = \sign \left( \popcount_j \left( \XNOR \left( W^{[k]}_{ij}, a^{[k-1]}_j \right) \right) -  \mu^{[k]}_i  \right),    
\label{eq:activation_binarized}
\end{equation}

where $\sign$ is the sign function.

Ordinarily, in binarized neural network, 
%whether it is a convolutional or a fully connected neural network, 
the first layer input $X$  is not binarized. 
The implementation of operations for computing the first layer activations $a^{[1]}$ is therefore more complex than the basic $\XNOR$ and $\popcount$ operations:

\begin{equation}
a^{[1]}_i = \sign  \left( \sum_{j}{W^{[1]}_{ij} \cdot X_j}  - \mu^{[1]}_i  \right).    
\label{eq:activation_layer1}
\end{equation}

Additionally, the thresholding operation is not performed on the last layer of the neural network.
Instead, for the last layer, we identify the neuron with the maximum $\popcount$ value 
(i.e. the  $\argmax$ of the last layer neurons), 
which gives the output of the neural network.
The whole  inference process of a conventional BNN  is described with vectorized notations in Algorithm~\ref{alg:algorithm1}.

The performance of BNNs is quite impressive.
{
 A fully-connected BNN with two hidden layers of 1024 neurons, and the use of dropout during training \cite{srivastava2014dropout}  obtains a $1.8\%$  error rate on the test dataset of the canonical MNIST handwritten digits task \cite{lecun1998gradient}, 
with 
300 epochs. 
In comparison,  a conventional neural network with no binarization and $\tanh$ activation function, and the same architecture and number of neurons, obtains a $1.5\%$ test error rate  after 300 epochs.} 
Similarly, on more complex datasets such as CIFAR-10 or ImageNet, near-equivalent performed is obtained by BNNs and conventional neural networks \cite{courbariaux2016binarized,rastegari2016xnor, lin2017towards}.
The  low memory requirements of BNNs (one bit by synapse), as well as the fact that they do not require any multiplication, makes them extremely adapted for inference hardware \cite{nurvitadhi2016accelerating,sun2018xnor,tang2017binary,yu2018neuro}.

The  training process of BNNs is reminded in  Appendix~\ref{sec:appendix_training}. 
Unlike inference, the training process requires real valued weights and real arithmetic: 
{training BNNs is not easier than in a conventional neural network}. 
Therefore, a natural vision is to train BNNs on standard GPUs, and to use specialized ultra-efficient hardware only for inference.

\begin{algorithm}[ht]
\caption{Conventional BNN Inference Model }
\label{alg:algorithm1}
\begin{algorithmic}[]
\REQUIRE{input vector  $X$, trained weight matrices  $ W $ and threshold vectors $\mu$ }
\ENSURE{predicted output}
\STATE {\textbf{1. Non binary first layer:}}
%\STATE $z^{[1]}_0 = 0$ 
%\FOR{$t = 1 \to T$}
\STATE $z^{[1]} \leftarrow  W^{[1]} \cdot X$
%\ENDFOR
\STATE $a^{[1]} \leftarrow \sign(z^{[1]} - \mu^{[1]})$
\STATE {\textbf{2. Remaining layers:}}
\FOR{$k = 2 \to L$}
  \STATE $z^{[k]} \leftarrow \popcount( \XNOR( W^{[k]}, a^{[k-1]}))$ 
  \STATE $a^{[k]} \leftarrow \sign(z^{[k]} - \mu^{[k]})$
  %\IF {$(k < L)$} $a^{[k]} \leftarrow sign(z^{[k]} - \mu^{[k]})$ 
  %\ELSE \STATE $a^{[k]} \leftarrow max(z^{[k]} - \mu^{[k]}) $ 
  %\ENDIF
\ENDFOR
  \STATE $z^{[L]} \leftarrow \popcount( \XNOR( W^{[L]}, a^{[L-1]}))$ 
%  \STATE $a^{[L]} \leftarrow max(z^{[L]} - \mu^{[L]}) $
 \STATE $output \leftarrow \argmax(z^{[L]} - \mu^{[L]}) $ 
\end{algorithmic}
\end{algorithm}

In this work, we investigate how the first layer can be approximated by a stochastic input to decrease computing resources. 
This approach could also allow  processing stochastic data for near sensor computing, which is a way to reduce considerably data transfer between sensors and data process.
In addition, due to the possibility of implementing binarization from the first layer, the model can be completely generic with exactly the same architecture over the layers and allows reducing chip area.

\subsection{Stochastic Computing}

Stochastic computing is an approximate computing paradigm, known since the early days of computing \cite{gaines1969stochastic,alaghi2013survey}. 
{Nevertheless, hardware engineers have not exploited this computing scheme for processor design,as 
it  requires applications that can be easily mapped with approximate computing. }
The principle is based on encoding {all  data} as probabilities,  represented as a temporal stochastic bitstreams: 
the number of ones among the bitstream represents the encoded probability. 
The main advantage of this {encoding scheme} is that  mathematical functions can be easily approximated with simple logic gates. For instance a product is then implemented with a single AND gate, and a weighted adder can be implemented with a multiplexer gate \cite{alaghi2013survey}. 
Many arithmetic operations are therefore easy to implement with low power
and small footprint characteristic. 
Despite these benefits, stochastic computing holds drawbacks: 
its limitation to low precision arithmetics, 
and the need to generate random bits. Random number generation can be a major part of the energy consumption, and, moreover, the generated random bits need to be uncorrelated. 

Random bits have also found applications in the field of neural networks. 
The most widely used neural networks that intrinsically exploit stochasticity are the restricted Boltzmann machine, where each neuron is binary valued with a probability that depends on the previous layer neurons states \cite{hinton2006fast}. 
An alternative technique to exploit stochasticity in neural networks  is to approximate standard neural network architecture with stochastic computing. 
This approach as been proposed as early as the 1990's \cite{bade1994fpga}, and is currently being revisited in modern deep neural networks  \cite{ardakani2017vlsi,ren2017sc,canals2016new}. 
These works have shown promising results in terms of area and energy consumption. Typically, the largest challenge is the implementation of the non-linear activation function within the stochastic computing framework.

In this article, we suggest that stochastic computing is particularly adapted to the case of binarized neural network,
as they work so naturally with bitstreams, and as the activation function is replaced by a simple thresholding operation.

%%%%%%%%%%%%%%%%%%%%%%%%%%%%%%%%%%%%%%%%%%%%%%%%%%%%%%%%%%%%%%%%%
%%%%%%%%%%%%%%%%%%%%%%%%%%%%%%%%%%%%%%%%%%%%%%%%%%%%%%%%%%%%%%%%%

\Figure [ht](topskip=0pt, botskip=0pt, midskip=0pt) [scale = 0.25]{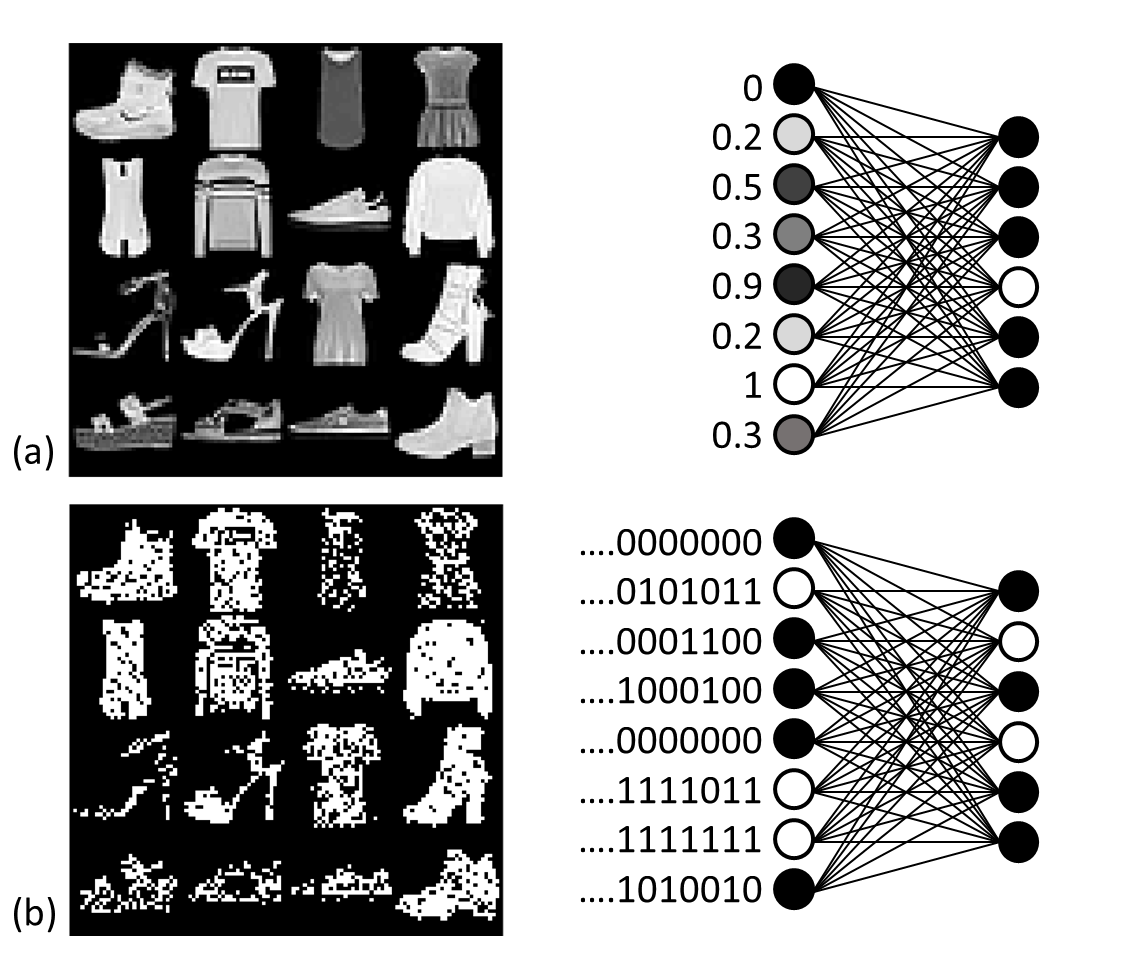}
{(a) In a conventional BNN, the first layer is not binarized. Grayscale input images are presented to the neural network. { (b) In a stochastic computing-based BNN,   binarized images are generated stochastically based on a grayscale image.} Several binarized versions of the same original image can be presented sequentially to the neural network, following the basic principle of stochastic computing.  \label{fig:stochprinciple}}

\section{Stochastic Computing-Based Binarized Neural Network}
\label{sec:network}

To evaluate {the stochastic computing approach}, we use the Fashion-MNIST dataset, which {has} the same format as MNIST, but presents  grayscale images of fashion items \cite{xiao2017fashion}, and constitutes a harder task. 
The canonical MNIST dataset would not be appropriate for this study, as it consists in images that are mostly black and white.
As in the MNIST dataset, each image in Fashion-MNIST has 28x28 pixels, and  can be classified within ten classes. The dataset contains  $60,000$ training examples, $10,000$ test examples. 
Conventional BNNs (non-binarized first layer and no use of  stochastic computing), perform very well on this task.
With a fully connected  BNN with first layer coded with eight bit fixed point real numbers, with  two hidden layers of 1024 neurons  each  and dropout, { a classification accuracy of $90\%$ can be obtained after 300 epochs}. 
This result is comparable with the test accuracy of $91\%$ obtained by a conventional real-valued neural network with  the same architecture.

\Figure [ht](topskip=0pt, botskip=0pt, midskip=0pt) [scale = 0.35]{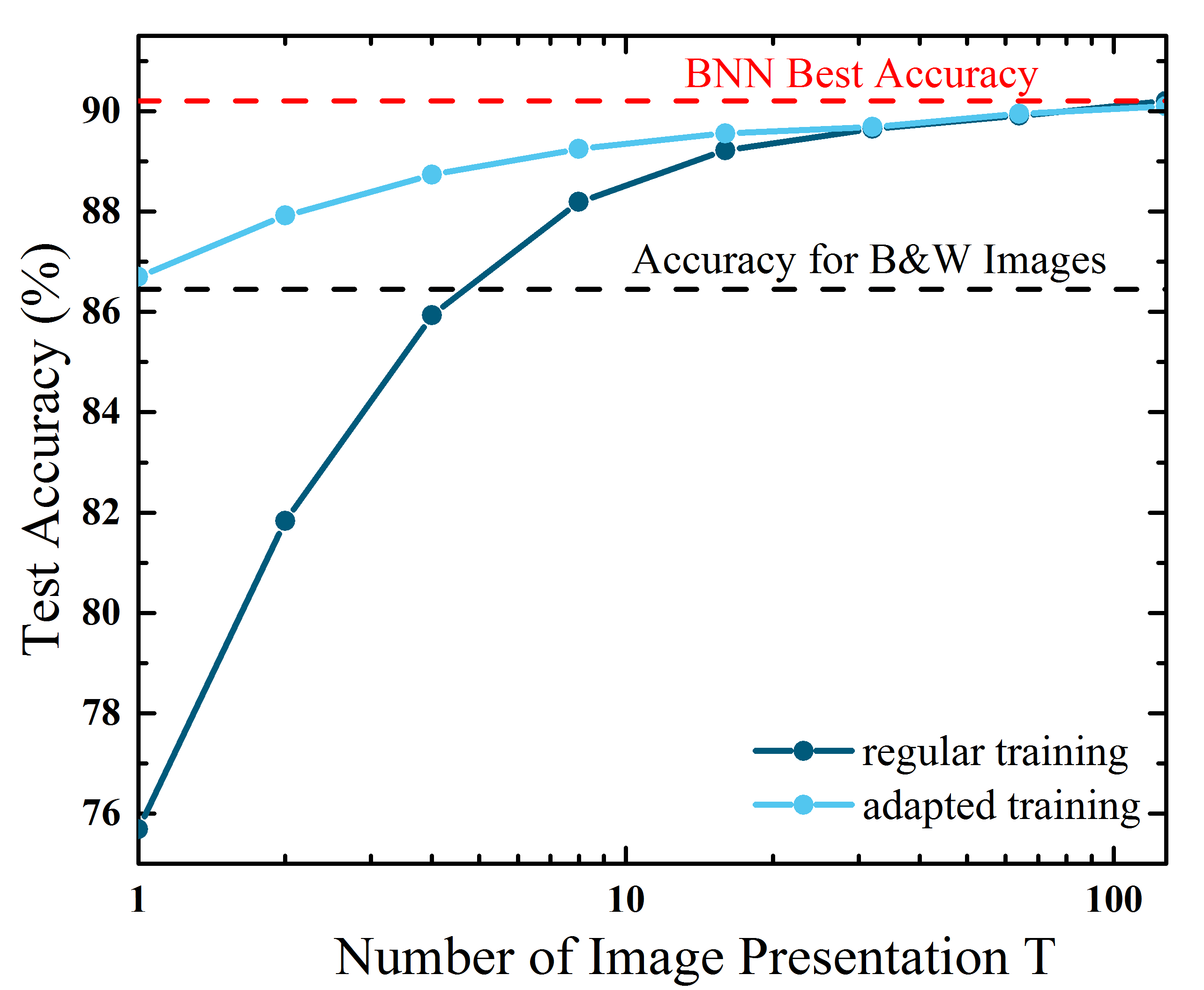}
{Accuracy on the Fashion MINIST classification task as function of the number of stochastic image presented for the two training methods. Navy blue curve: training of the neural network with grayscale images. Light blue curve: training  with presentation of stochastic binarized images. Dashed black line: accuracy when training with a black and white image (i.e. pixels with a value greater than 0.5 are white and pixels that are smaller are black). { Dashed red line: best accuracy when  the binarized neural network is trained on Fashion-MNIST classification task with grayscale images}. 300 training epochs were used.
\label{fig:accuracy}}

\subsection{Stochastic Computing with Regular Training Procedure}
\label{subsec:firstalyer}

\begin{algorithm}[ht]
\caption{Stochastic Computing BNN with Binarized First Layer}
\label{alg:algorithm1binary}
\begin{algorithmic}[]
\REQUIRE{$ X_t $ vectors: T stochastic binary versions of non-binary input X, trained weight matrices  $ W $ and threshold vectors $\mu$ }
\ENSURE{predicted output}
\STATE {\textbf{1. Stochastic and binarized first layer}}
\STATE $z^{[1]} \leftarrow 0$ 
\FOR{$t = 1 \to T$}
\STATE $z^{[1]} \leftarrow z^{[1]} + \popcount( \XNOR( W^{[1]}, X_t))$
\ENDFOR
\STATE $a^{[1]} \leftarrow \sign(z^{[1]} - T \mu^{[1]})$
\STATE {\textbf{2. Remaining layers}}
\FOR{$k = 2 \to L$}
  \STATE $z^{[k]} \leftarrow \popcount( \XNOR( W^{[k]}, a^{[k-1]}))$ 
  \STATE $a^{[k]} \leftarrow \sign(z^{[k]} - \mu^{[k]})$
  %\IF {$(k < L)$} $a^{[k]} \leftarrow sign(z^{[k]} - \mu^{[k]})$ 
  %\ELSE \STATE $a^{[k]} \leftarrow max(z^{[k]} - \mu^{[k]}) $ 
  %\ENDIF
\ENDFOR
  \STATE $z^{[L]} \leftarrow \popcount( \XNOR( W^{[L]}, a^{[L-1]}))$ 
  \STATE $output \leftarrow \argmax(z^{[L]} - \mu^{[L]}) $ 

\end{algorithmic}
\end{algorithm}

{
A first approach to design a stochastic computing BNN is to reuse the synaptic weights of a conventional BNN, trained with grayscale picture.}
However, in the inference phase, we approximate  the  computation of the first layer by using stochastic images presentation instead of grayscale images.
The full inference algorithm is presented, in vectorized form, in Algorithm~\ref{alg:algorithm1binary}.
An input $X$ is transformed into binarized
 stochastic inputs $X_t$ by taking the value of each grayscale pixel (between zero and one) as the probability for the corresponding pixel in the stochastic input to be one.
Then,  {the networks computes} $\popcount( XNOR( W^{[1]}, X_t)) - \mu^{[1]}$, and  sums the result of this computation over a number $T$ of stochastic versions of the input $X_t$. 
Finally, the output of the layer is thresholded to obtain a binary value, and the rest of the neural network is computed in one pass in a fully binarized fashion.

The quality of the results depends on the number of  image presentation $T$. 
In Fig.~\ref{fig:accuracy}, the navy blue curve shows the network test error as a function of $T$.
We can see that after 100 stochastic image presentation, the accuracy is nearly equivalent to  the use of  grayscale images. 
With eight image presentation, the test accuracy is reduced to $88\%$ instead of $90.1\%$. 
With a single presentation, the test accuracy is only $76\%$ 

\subsection{Adapted Training Procedure}
\label{sec:adaptedtrainig}

We now try a second strategy, where we train the neural network with binarized stochastic image presentation instead of grayscale images. 
To do this, during training, we use the conventional BNN training technique of Appendix~\ref{sec:appendix_training}, but instead of using the normal grayscale Fashion-MNIST images, we use stochastic binarized ones, with the same number of presentation $T$ as will be used during inference.
The inference technique then remains identical to the one described in section~\ref{subsec:firstalyer}.
% backpropagate the error through all layers without considering multiple presented stochastic images. 
%For inference, we use the exact same technique as in the section~\ref{subsec:firstalyer}, which
% means that after the stochastic training, we now present multiple stochastic images of the same image. 
In Fig.~\ref{fig:accuracy}, in cyan color, we plotted the test error rate as a function of the number of presentation of the same image with this scheme. 
We see that the test accuracy is equivalent to the one obtained with grayscale images for high numbers of image presentation. On the other hand, with  few stochastic presentation (one to five), the adapted input training technique allows reaching a quite high accuracy. 
If a single presentation is used at inference time, the network test accuracy is $86\%$.
This test accuracy is equivalent to the one obtained when training a BNN with non-stochastic black and white versions of the Fashion-MNIST dataset 
(dashed black line in Fig.~\ref{fig:accuracy}). 
If three image presentation are used, the network test accuracy increases to {$88.7\%$}.

These results show that when using the stochastic computing version of BNN, the adapted training procedure should be used.
%both training strategies (training the network with grayscale images and with binarized images) have value, but should not be used in the same contexts.
%When favoring energy consumption (one or few presentations of the images during inference), training with binarized images is more beneficial.
%When favoring network accuracy (many presentations of the images during inference), training with grayscale images is more beneficial.

\subsection{Choice of the Accumulation Layer for Stochastic Samples}
\label{subsec:accumul}

\Figure [ht](topskip=0pt, botskip=0pt, midskip=0pt) [scale = 0.35]{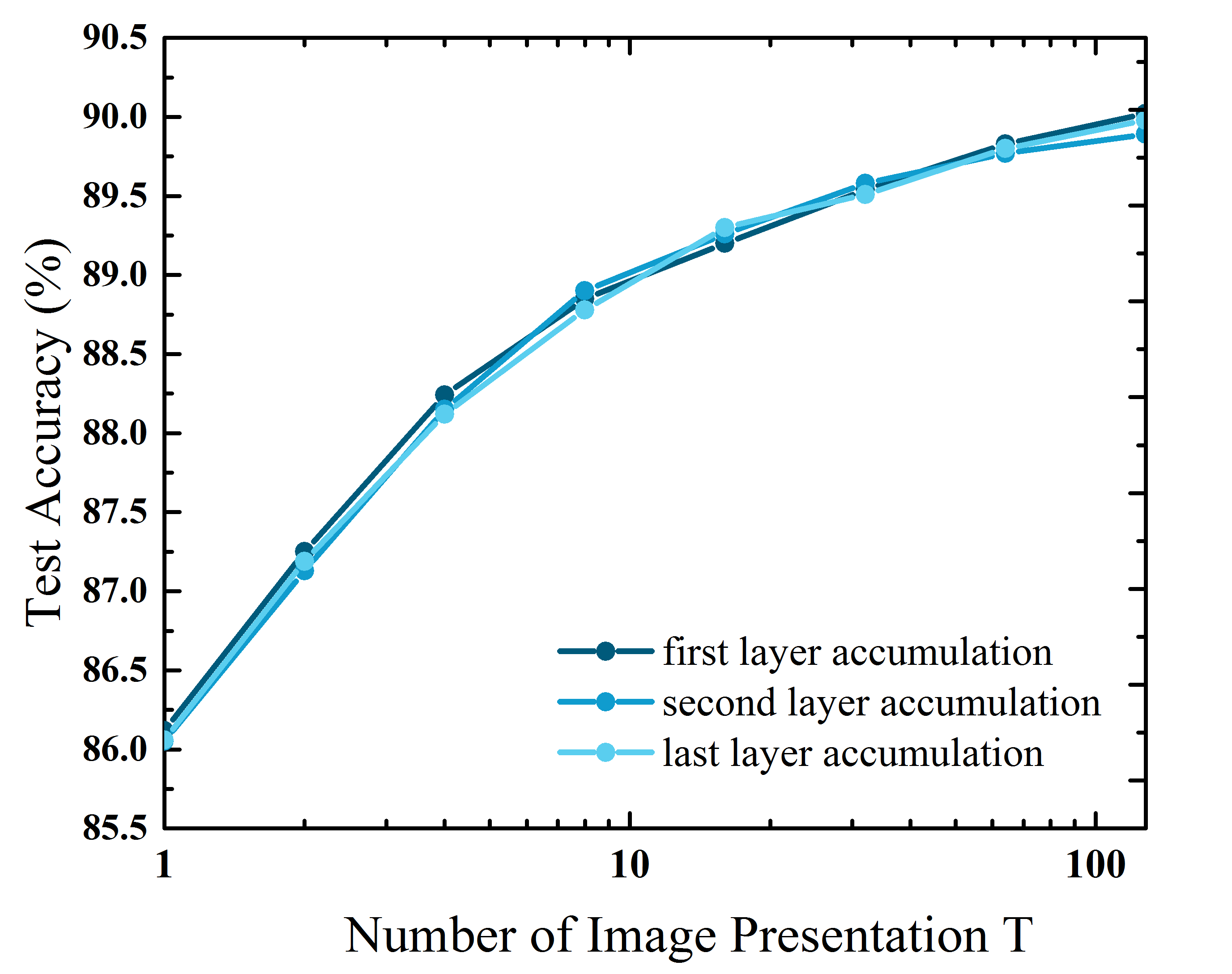}
{Accuracy on the Fashion MINIST classification task as function of the number of stochastic image presentation presented { when  the popcount value ias accumulated at different level of the network. The training was done with grayscale images and }300 training epochs were used.  \label{fig:accumulation}
}

Until now, at inference time, we have accumulated the outputs of the first layer over  several presentations of the same image, then propagated the binarized output of the first layer to the other layers.
An alternative strategy can be to perform the accumulation over the  realizations of the input images at another layer. If the accumulation is done at the last layer,
this procedure corresponds to using stochastic computing in the whole depth of the neural network. 

Fig.~\ref{fig:accumulation} presents  the test accuracy of the neural network on the Fashion-MNIST dataset, as a function of the number of presented realizations of the input images, for the different accumulation strategy, in networks trained with the adapted training strategy.
{This Figure shows that that the different accumulation strategy lead to equivalent accuracy, consistently with the principles of stochastic computing. }
{The strategy  of accumulation at the first layer is retained for the rest of the paper, as it allows for the minimum energy consumption.}

\subsection{ {Extension to the CIFAR-10 Dataset} }
\label{subsec:CIFAR10}

\Figure [ht](topskip=0pt, botskip=0pt, midskip=0pt) [scale = 0.35]{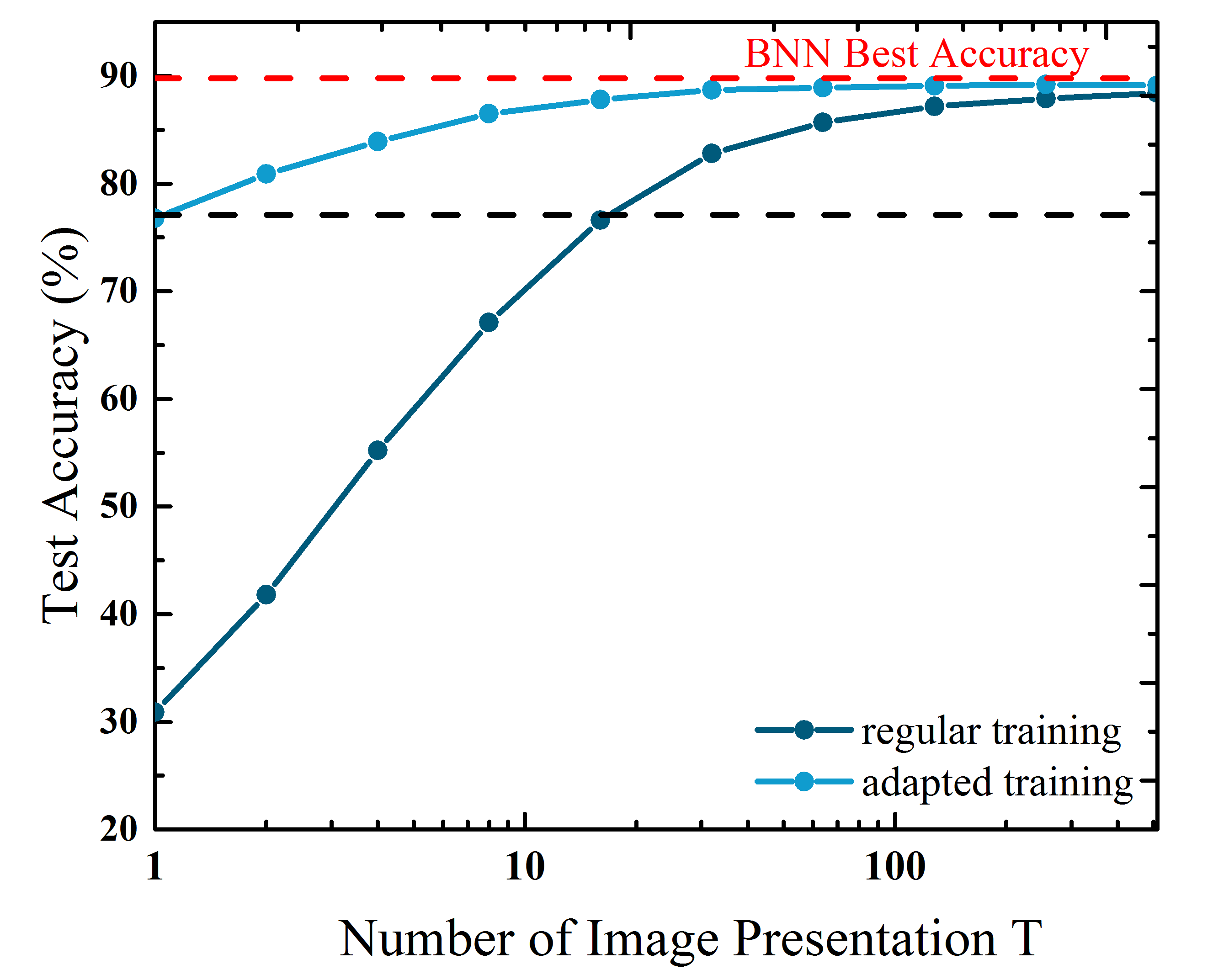}
{{Accuracy on the CIFAR-10 classification task as function of the number of stochastic image presented for the two training methods. Navy blue curve: training of the neural network with color images. Light blue curve: training  with presentation of stochastic binarized images. Dashed black line: accuracy when training with a binarized color image (i.e. RGB values with a value greater than 0.5 are white and pixels that are smaller are black).  Dashed red line: best accuracy when  the binarized neural network is trained on CIFAR-10 classification task with full color images. 2000 training epochs were used.
}
\label{fig:accuracy_cifar_bin}}

\Figure [ht](topskip=0pt, botskip=0pt, midskip=0pt) [scale = 0.35]{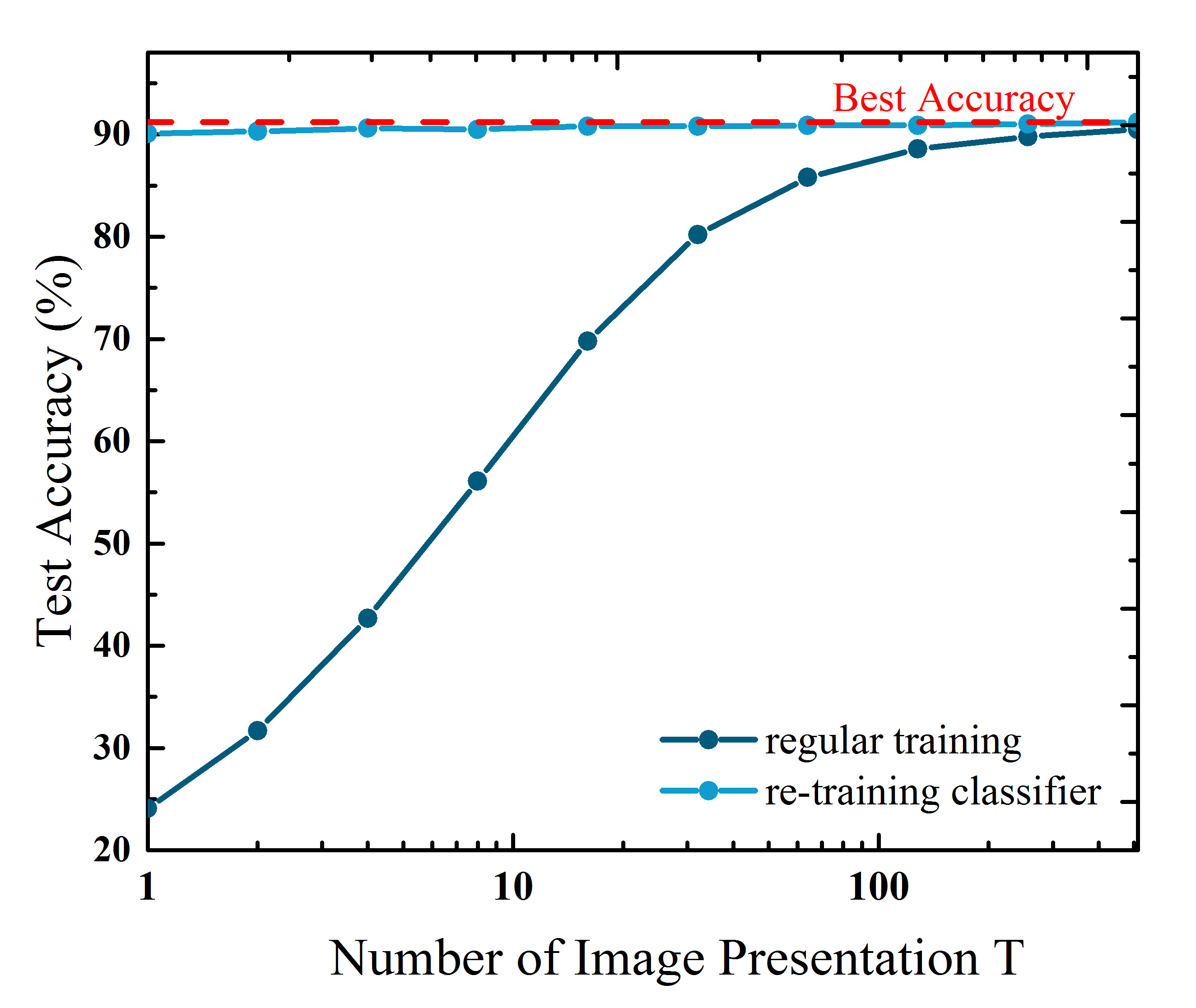}
{{Accuracy on the CIFAR-10 classification task, but the stochastic computing approach is implemented at the end of the convolutional layers.  Navy blue curve: training of the neural network in a conventional fashion. Light blue curve: classifier part of the neural network retrained with stochastic versions of the output of the convolutional layers. Dashed red line: best accuracy when  the binarized neural network is trained on CIFAR-10 classification task with full color images. 2000 training epochs were used.}
\label{fig:accuracy_cifar_bin_classifier}}

{
We now apply this strategy to the more advanced CIFAR-10 dataset.
We use a convolutional neural network with six convolutional layers, with kernel size of three by three and a stride of one (number of filters 384, 384, 384, 768, 768 and 1536) and three fully connected layers (number of neurons 1024, 1024 and 10). 
Training is done in the same conditions as the Fashion-MNIST case, using dropout and Adam optimizer, and the pytorch deep learning framework.
In the stochastic computing BNN, CIFAR-10 images are presented with binarized channel: 
each RGB channel pixel presents a value of zero or one. This value is chosen randomly with a probability equal to the RGB value of the corresponding pixel of the image.
Accumulation of stochastic realization is realized at the first layer, as described in section~\ref{subsec:accumul}.
}

{
Fig.~\ref{fig:accuracy_cifar_bin} shows that the results on CIFAR-10 are very similar to the ones on Fashion-MNIST (Fig.~\ref{fig:accuracy}). 
It present results obtained using the weights trained with full color images, and  weights obtained with the adapted training approach. In both cases, the stochastic BNN results approach regular BNN results when the number of presentation $T$ of stochastic images is increased. The adapted training nevertheless gives highly superior results and should be preferred.
This highlights that the stochastic BNN approach can be applicable to more complicated tasks than Fashion-MNIST.
}

{
We now consider a variation of this scheme, a partially binarized neural network.
Fully connected layers of neural networks are particularly adapted for in-memory  BNN implementation \cite{yu2018neuro,bocquet2018memory}, as these layers involve large quantities of memories. 
Convolutional layers are less memory intensive, and thus benefit less from binarization, while requiring increasing the number of channels when binarized \cite{courbariaux2016binarized}.
In a hardware implementation, it can therefore be attractive to binarize only the classifier (fully connected) layers. In that case, the input of the classifier is real, and is processed with the stochastic BNN approach. 
This is also of special interest as the first fully connected layer in a convolutional neural network is usually the layer that involves the highest number of additions, and can therefore benefit significantly in a hardware to be implemented with the stochastic approach. 
}

{
We consider a neural network with the same architecture as the fully binarized one,  a
reduced number of filters (128, 128, 128, 256, 256 and 512) and the same number of neurons in the fully connected layers (1024, 1024 and 10). Without the stochastic approach, this neural network has the same CIFAR-10 recognition rate than the fully binarized one ($90\%$).
Fig.~\ref{fig:accuracy_cifar_bin_classifier} shows the results of the stochastic BNN with this approach.
If the same weights are used than in a non stochastic BNN, the results look  similar to the fully binarized approach of Fig.~\ref{fig:accuracy_cifar_bin}. On the other hand, if the classifier weights are retrained with the stochastic binarized inputs to the classifier, the stochastic results are very impressive. 
Even with a single image presentation $T$, the network approaches the performance of the non stochastic network.
The stochastic BNN approach therefore appears especially effective in this situation.
}

\section{Hardware Implementation of Stochastic Computing-Based Binarized Neural Network}
\label{sec:hardware}

In order to investigate the potential of the stochastic BNN approach, we designed a digital ASIC version of it using standard integrated circuit design tools. 
The architecture, presented in Fig.~\ref{fig:archi}, 
 allows performing the inference of a fully connected binary neural network of any size (up to 1024 neurons for each layer). 
The only parameter constrained by the hardware design is  the number of weights that can be stored.

\subsection{Design of the Architecture}

\Figure [ht](topskip=0pt, botskip=0pt, midskip=0pt) [scale = 0.17]{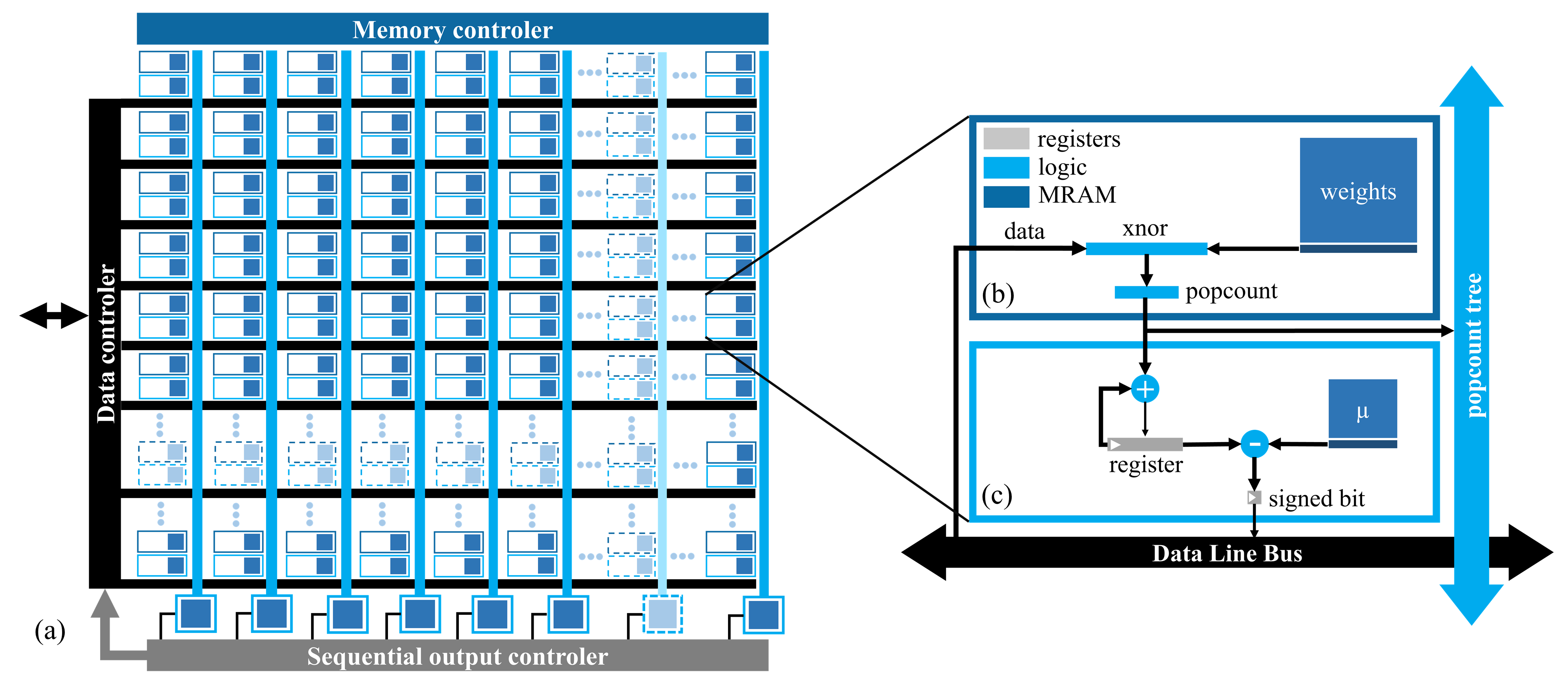}
{Design of an MRAM based fully connected binarized neural network, computing both parallel and serial. (a) Full architecture with $32 \times 32$ repeated cells. Each cell (b-c) behaves as a neuron if the input is sequential, or each column behaves as a neuron if the input is parallel. \label{fig:archi}}

Our architecture  is inspired by the works of \cite{ando2017brein}, 
with Static RAMs replaced by Spin Torque MRAM \cite{shum2017cmos},
and adaptation to stochastic computing. 
This architecture aims at performing inference on binarized neuronal networks with minimal energy consumption. 
To achieve this goal, it  brings memory and computation as close as possible, to limit energy consumption related to data transfer. 
Such an architecture takes special interest with the emergence of new non-volatile memory components such as Spin Torque MRAM, which can be integrated within the CMOS manufacturing process, and which we consider here.

The architecture  is described in detail in Appendix~\ref{sec:appendix_system}, and  {can compute following  
a parallel or a sequential structure}.
{The full design is made by a basic cell repeated 32x32 times (Fig.~\ref{fig:archi} (b-c)) that can perform both sequential or parallel calculation. 
It includes a 2~kbits memory array to stores weights, as well as XNOR gates and popcount logic.}

We designed this system using the design kit of a commercial 28~nanometer technology. 
{Digital circuits were described  in synthesizable  SystemVerilog description.}
MRAM memory arrays are modeled in a behavioral fashion, and their characteristics (area, energy consumption) are inspired by \cite{chun2013scaling}.
The system was synthesized to estimate its area and energy consumption.
For energy consumption, we employed Value Change Dumps extracted from a Fashion-MNIST inference task, and estimated it using the Cadence Encounter tool.

\subsection{Energy Consumption and Area Results}

\Figure [ht](topskip=0pt, botskip=0pt, midskip=0pt) [scale = 0.35]{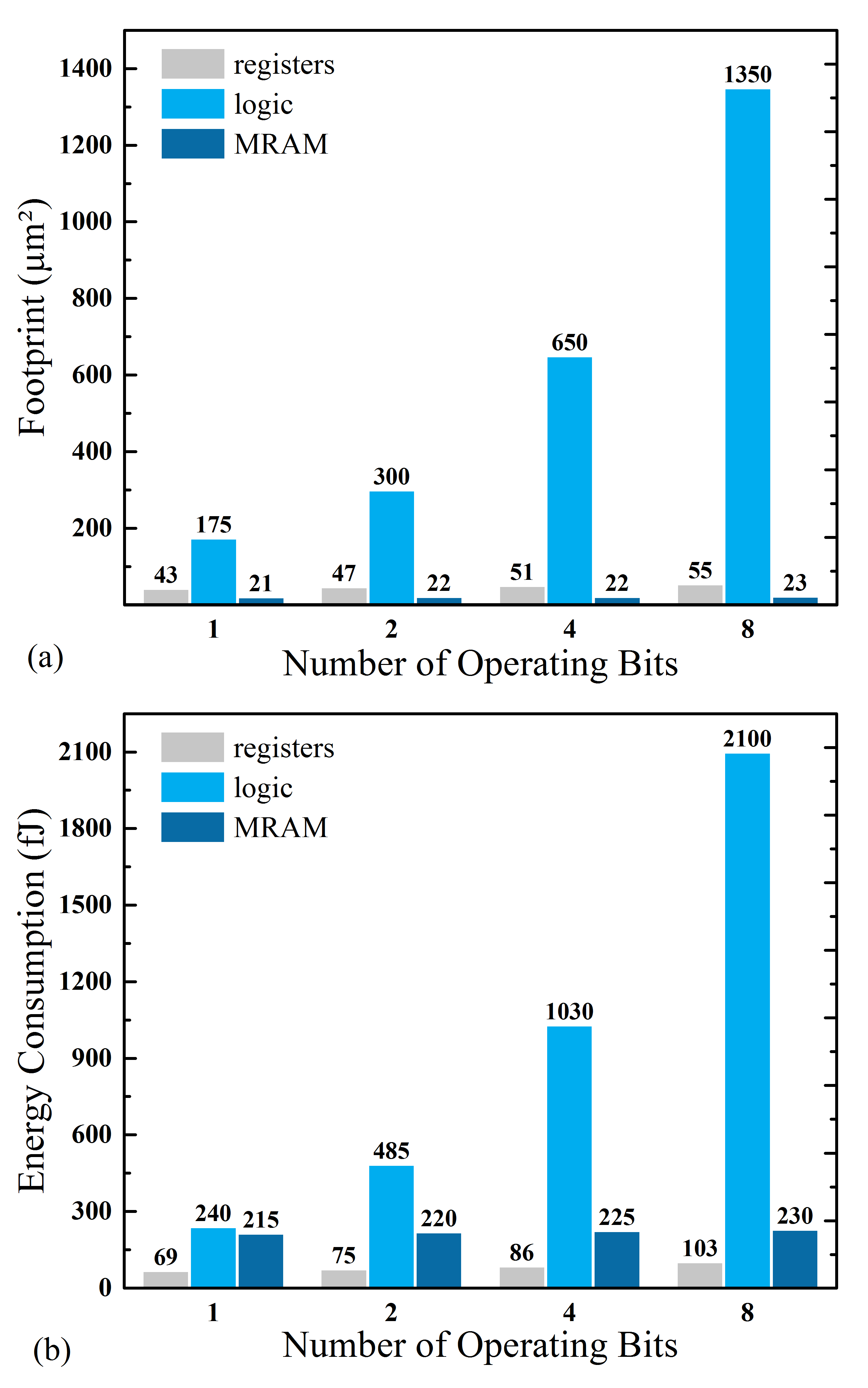}
{(a) Area of the basic cell (Fig. ~\ref{fig:archi} (b-c)) of our ASIC architecture, implemented in a 28~nm CMOS technology,
as function of the number of operating bit for a fixed point binary architecture. One-bit corresponds to our stochastic fully binarized architecture. (b) Corresponding energy consumption, per clock cycle. \label{fig:area}}

Fig.~\ref{fig:area}(a) shows the area of a basic cell of our architecture  (Fig.~\ref{fig:archi}(b-c)), in the case of binary input (one operating bit), and in situations where the input is coded in Fixed Point representation (two, four and eight operating bit), as is required in the first layer of a conventional BNN.
{
This Figure separates the area used by registers, logic and MRAM.
A cell with binary input uses six times less area than a cell designed for eight bit input. }
Interestingly, the difference is mostly due to the $\popcount$ circuits, which need more depth when the input is non-binary.
Similarly, as seen in Fig.~\ref{fig:area}(b), a cell with binary input uses $4.5$ times less energy per cycle than the corresponding one with eight bits input. 
Again, the difference is  mostly due to the $\popcount$ circuits. 

\Figure [ht](topskip=0pt, botskip=0pt, midskip=0pt) [scale = 0.35]{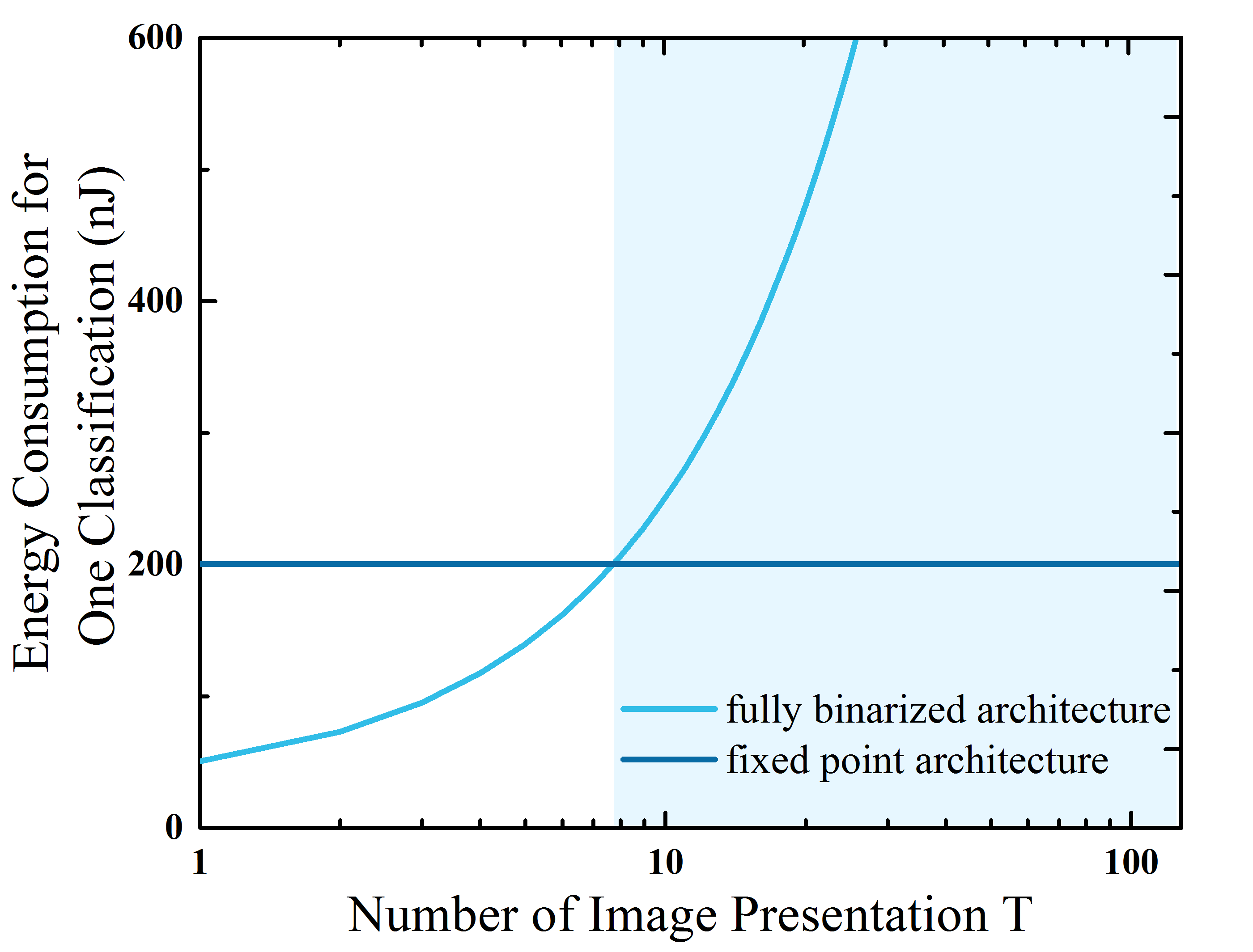}
{Energy consumption of the full Fashion-MNIST classifier systems, for the classification of one image. Light blue: stochastic  fully binarized binary architecture. Navy blue:  Conventional BNN with non binary (8 bit fixed point) first layers. The neural networks have two layers with 1024 neurons each. The light blue area indicate the regime where the non-binary first layer is more energy efficient thant the fully binarized system.\label{fig:energy} }

{The savings in terms of area transfer directly at the system level. 
We  now consider the whole neural network used for Fashion-MNIST classification throughout section \ref{sec:network}.}
Using our architecture, a full BNN with eight bit first layer  occupies $1.95\ mm^2$, while the BNN with stochastic binarized first layer occupies $0.73\ mm^2$, a $62\%$ saving in area. 
These area values were extracted from a system designed  for a  $T$ value of eight.

Fig.~\ref{fig:energy} plots the energy consumption for recognizing an image with our ASIC architecture, as a function of the number of presented stochastic images.
This is compared with the energy cost of the same architecture, but using a non stochastic first layer, with eight bit input.
We see that the system with stochastic first layer is more energy efficient than the system with non-binary  first layer if less than eight presentation are used. 

The previous curves do not include the  cost of random bit generation.
If we use a simple eight-bit Linear Feedback Shift Register (LFSR) pseudo random number generator, the added energy is $0.52nJ/cycle$,  and the added area is $48,000\ \mu m^2$. Both are therefore negligible.
It has also been shown that Spin Torque MRAM technology can be adapted to provide very low energy true random numbers \cite{vodenicarevic2017low}.
If such a technology was used, based on the numbers of  \cite{vodenicarevic2017low}, the energy cost of random bit generation would be  $0.125nJ/cycle$, and the area much smaller than LFSR. 
The energy cost of random number generation is therefore negligible with regards to the consumption of the system seen in Fig.~\ref{fig:energy}.

{These energy numbers are very attractive with regards to non binarized implementations at equivalent recognition rate. Non binarized neural networks require less neurons and synapses than BNNs to achieve equivalent recognition rate. For example, to match the performance obtained in Fig.~\ref{fig:accuracy} on Fashion-MNIST with three image presentations ($T=3$), one only needs a non-binarized neural network with eight-bit synapses with two layers of 500 neurons, while the BNN needs 1024 neurons per layer.
However, in an ASIC, the non binarized neural network requires energy-hungry 8-bits multiplications and addition ($0.3~pJ$ and $0.04~pJ$ per operation in our $28~nm$ technology). Taking into account only these arithmetic operations, the energy consumption is $220~nJ$ for recognizing a Fashion-MNIST image with the same accuracy as the stochastic BNN with three image presentations. This stochastic BNN requires only $90~nJ$ (Fig.~\ref{fig:energy}), taking into account the whole system.
}

{
As a conclusion, this works highlights that the stochastic computing approach is  attractive in terms of area occupancy.} 
In terms of energy efficiency, it is very attractive  if a relatively small number of presentation is used ($T<8$).
Therefore, it appears preferable to rely on the stochastic training approach seen in section \ref{sec:adaptedtrainig}, and to use few stochastic image presentation for inference.
For example, if three image presentation are used,  a factor $2.1$ can be saved on the energy consumption on Fashion-MNIST, with a reduction of {$1.4\%$} of test accuracy with regards to the best accuracy obtained by a BNN (dashed red line in Fig.~\ref{fig:accuracy}). 
It should be noticed that the benefits of stochastic computing would be reduced on very deep neural networks, where the first layer plays a smaller role.  Our approach is therefore the most promising for Internet-of-Things or sensor networks applications, where relatively small neural networks can provide sufficient intelligence, but circuit cost and energy consumption are the most critical issues. {On deep neural networks, nevertheless, the approach of implementing only the classifier with a stochastic BNN, as mentioned in section~\ref{subsec:CIFAR10}, can be of high interest.}

%%%%%%%%%%%%%%%%%%%%%%%%%%%%%%%%%%%%%%%%%%%%%%%%%%%%%%%%%%%%%%%%%
%%%%%%%%%%%%%%%%%%%%%%%%%%%%%%%%%%%%%%%%%%%%%%%%%%%%%%%%%%%%%%%%%

\section{Conclusion}

In this work, we presented a stochastic computing approach to Binarized Neural Networks. 
This allows implementing them in an entirely binarized fashion, whereas in conventional BNNs, the first layer is not binary.
We showed that the stochastic computing approach can reach recognition results similar to the conventional approach.
We identified that for highest accuracy, the neural network should not be trained with  {regular} images as conventional BNNs:
it it is more beneficial to train stochastic BNNs with stochastic binarized images, using the same number of image presentation as will be used during inference.
{
The design of a full BNN ASIC relying on in-memory computing,  then highlighted the benefits of BNNs in terms of area and energy consumption.}
Stochastic BNNs allow using the same compact architecture for all layers, which leads to strong benefits in terms of area  ($62\%$ reduction in the case of Fashion-MNIST classification).
In terms of energy, the benefits can be very strong if we accept a slight reduction in classification accuracy. 
For example, on Fashion-MNIST classification, we can reduce the energy consumption by a factor {$2.1$}, with a decrease of {$1.4\%$} in classification accuracy. 

These results highlight the high potential of BNNs for implementing compact and energy efficient in-memory neural networks, and the potential of stochastic approaches for hardware artificial intelligence.
{Future works should focus on the physical implementation of the proposed scheme, as well as the extension of the approach to other tasks than vision, such as medical tasks, where energy efficiency can be a particularly important concern.}

\appendices 
\section{Training Algorithm}
\label{sec:appendix_training}

Throughout the paper, neural networks are trained with the algorithm proposed by Courbariaux et al in \cite{courbariaux2016binarized}. 
This algorithm relies on two fundamental principles. First, the function $\Clip(x, -1,1)$ is used instead of the  $\sign$ function in the backpropagation phase, as it can be differentiated. Second, the binarized weights $W$ are not directly modified during the back propagation: their modification is done indirectly through the modification of the real weight $W_a$ associated with each synapse. 

{Our design includes
two modifications with regards to the work of  \cite{courbariaux2016binarized}.}
In the original paper, the multi-layer perceptron  trained on MNIST  consisted of hidden layers of binarized units, topped by L2-SVM output layer. Here, we used a $\softmax$ output layer.
Second,  the parameters $\gamma$  and  $\beta$ used for the batch normalization were not trained, and we used $\gamma=1$ and $\beta=0$ instead. 
The complete algorithm that we used is presented in Algorithm~\ref{alg:algorithmtrain}.

\begin{algorithm}[ht]
\caption{Conventional BNN training model}
\label{alg:algorithmtrain}

\begin{algorithmic}[]
\REQUIRE{training data : $ X_{train}$, targets output $y_{train}$, previous binarized and real weights $W$ and $W_a$, and previous threshold values $\mu$ }
\ENSURE{updated weights $W_{t+1}$ and $W_{a,t+1}$, updated BatchNorm parameters $\mu$  and  $\sigma$} 
\STATE \textbf{1. Forward propagation}
\FOR{$k = 1$ to $L$}
  \STATE $W^{[k]} \leftarrow \sign( W^{[k]}_a )$ 
  \STATE $z^{[k]} \leftarrow W^{[k]} \cdot a^{[k-1]}$
  \STATE $\widehat{z}^{[k]} \leftarrow \BatchNorm(z^{[k]},\mu^{[k]},\sigma^{[k]})$
  \IF {$(k < L)$} 
  \STATE $a^{[k]} \leftarrow \sign(\widehat{z}^{[k]})$
  \ELSE 
  \STATE $a^{[k]} \leftarrow \softmax(\widehat{z}^{[k]})$
  \ENDIF
\ENDFOR
\STATE Compute gradient of softmax cross entropy loss :
\begin{equation*}{} g_{a^{[L]}} = \dfrac{\partial C}{\partial a^{[L]}} = a^{[L]} - y \end{equation*}
\STATE \textbf{2. Backward propagation}
  \FOR{$k = L$ to $1$}
  \IF {$(k < L)$} 
  \STATE$g_{a^{[k]}} \leftarrow g_{a^{[k-1]}}\ \circ \ 1_{|a_{k<1}|} $
  \ENDIF
\STATE $ g_{\widehat{z}^{[k]}} \leftarrow \BackBatchNorm(g_{a^{[k]}},\widehat{z}^{[k]},\mu^{[k]},\sigma^{[k]}) $
\STATE $ g_{z^{[k]}} \leftarrow W^{[k]\ T}g_{\widehat{z}^{[k]}} $
\STATE $ g_{W_b^{[k]}} \leftarrow g_{\widehat{z}^{[k]}} \ a^T_{k-1}$
\ENDFOR
\STATE \textbf{3. Update parameters}
\FOR{$k = 1$ to $L$}
  \STATE $W^{[k]}_{a,t+1} \leftarrow \Clip(\UpdateAdam(W_{a,t+1}^{[k]},g_{W_b^{[k]}}),-1,1) $
  \STATE $(\mu^{[k]},\sigma^{[k]})_{t+1} \leftarrow \MovingAverage(\mu_B^{[k]},\sigma_B^{[k]})_{t} $ 
\ENDFOR

\end{algorithmic}
\end{algorithm}

\section{Description of the ASIC BNN Architecture }
\label{sec:appendix_system}
{
The architecture for hardware implementation of BNN inference is presented in Fig.~\ref{fig:archi}.}
The basic function of a BNN is  to compute $\popcount( \XNOR( W, X)) - \mu$. 
To perform this function, first, the system needs to perform the XNOR between the inputs $X$ and the weights $W$, stored in the Spin Torque MRAM memory blocks. 
Second, it needs to perform the $\popcount$ function, and then compare this value with a threshold. 

To achieve this goal, the architecture is made of basic cells (cell Fig.~\ref{fig:archi} (b-c)), composed of a 2~kbits memory array that store weights, 32 XNOR logic gates that perform the XNOR between the 32~bits weights and the 32~bits received data, a 32~bits to 5~bits popcount module compound of basic tree adders.  The basic cell is repeated 32x32 times.

The architecture can be operated with a ``parallel to sequential'' structure, or a ``sequential to parallel'' structure.
The sequential to parallel structure allows  dealing with long input sequence data, and outputs a limited parallel output data. 
By contrast, the parallel to sequential structure allows  dealing with limited parallel input data, and outputs long sequence data.
The basic cells of Fig.~\ref{fig:archi} (b-c)  can perform both, sequential or parallel calculation. 
The output of the popcount can be given to the sequential part of the cell or to the parallel part of the system that will perform the popcount through the whole column, with a ``popcount tree'' module shared with all the cells of the column. 
The sequential section of the cell that receive the popcount output will perform the full popcount operation sequentially by summing the popcount output using a register. 

{
To perform the activation function of the neuron, the system adds in each cell the threshold values $\mu$ in a memory array. }
The signed bit of the difference between the popcount value saved in the register and $\mu$ gave the activation value. 
The same operation is made with the output of the popcount tree shared along the column.

\FloatBarrier
\bibliographystyle{IEEEtran}
\bibliography{biblio}

\clearpage

\begin{IEEEbiography}[{\includegraphics[width=1in,height=1.25in,clip,keepaspectratio]{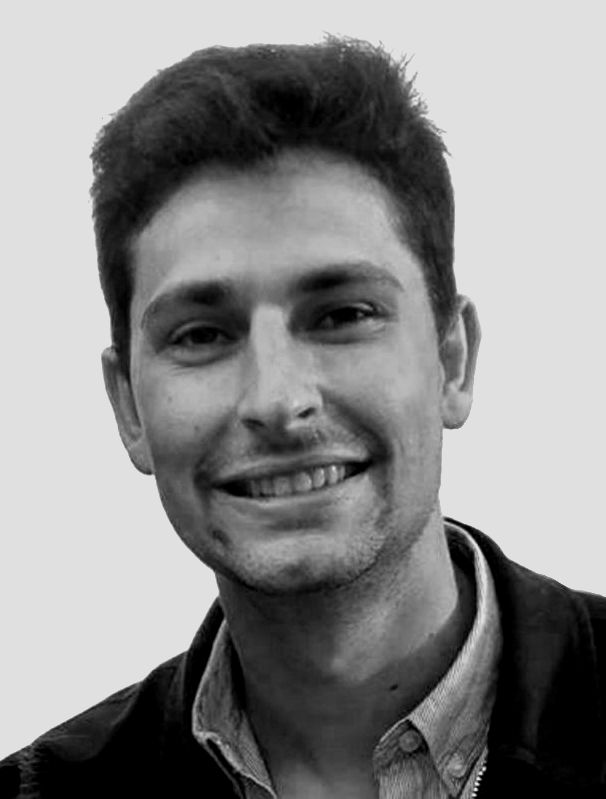}}]{Tifenn Hirtzlin} is a PhD student in Electrical Engineering at  Universit\'e Paris-Sud. He received the M.S. degree in Nanosciences and Electronics from the University Paris-Sud, France, in 2017.
His work focuses on designing intelligent memory-chip for low energy hardware data processing using bio-inspired concepts as probabilistic approach to brain function or more classical neural network approaches. 
\end{IEEEbiography}

\begin{IEEEbiography}[{\includegraphics[width=1in,height=1.25in,clip,keepaspectratio]{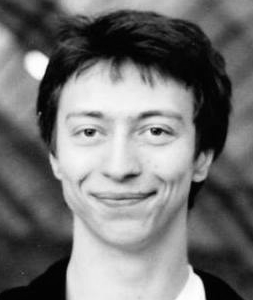}}]{Bogdan Penkovsky} is a postdoctoral CNRS researcher at Paris-Sud University. He received his M.S. degree in Applied Mathematics from the National University of Kyiv-Mohyla Academy, Ukraine, in 2013
and the Ph.D. degree in optics and photonics applied
to neuromorphic computing from the University of Burgundy - Franche-Comt\'e, France, in 2017.
His work is on intelligent, low energy hardware design for biomedical applications.
\end{IEEEbiography}

\begin{IEEEbiography}[{\includegraphics[width=1in,height=1.25in,clip,keepaspectratio]{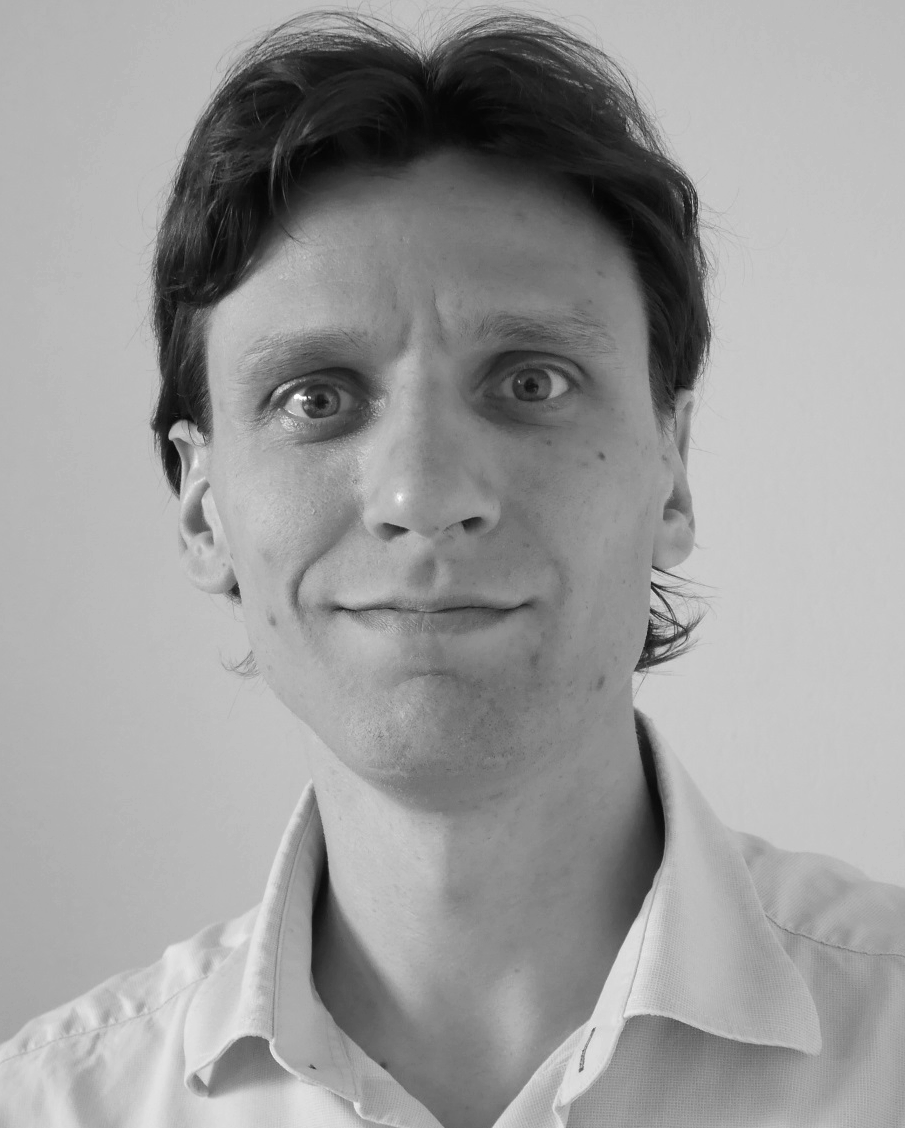}}]{Marc Bocquet} is an Associate Professor in  the Institute of Materials, Microelectronics and Nano-sciences of Provence, IM2NP at  Univerisity of Aix-Marseille. He received the M.S. in electrical engineering degree in 2006 and the Ph.D. degree in electrical engineering in 2009, both from the University of Grenoble, France. His research interests include memory model, memory design, characterization and reliability.
\end{IEEEbiography}

\begin{IEEEbiography}[{\includegraphics[width=1in,height=1.25in,clip,keepaspectratio]{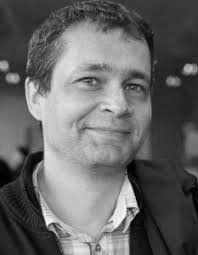}}]{Jacques-Olivier Klein}  (M'90) received the Ph.D. degree
from  Univ. Paris-Sud, France, in 1995. He is
currently Full Professor at Univ. Paris-Sud, where he
 focuses
on the architecture of circuits and systems based on
emerging nanodevices in the field of nanomagnetism
and bio-inspired nanoelectronics. In addition, he is lecturer at the Institut Universitaire de Technologie (IUT) of Cachan.
 He is author of more than one hundred
technical papers.
\end{IEEEbiography}

\begin{IEEEbiography}[{\includegraphics[width=1in,height=1.25in,clip,keepaspectratio]{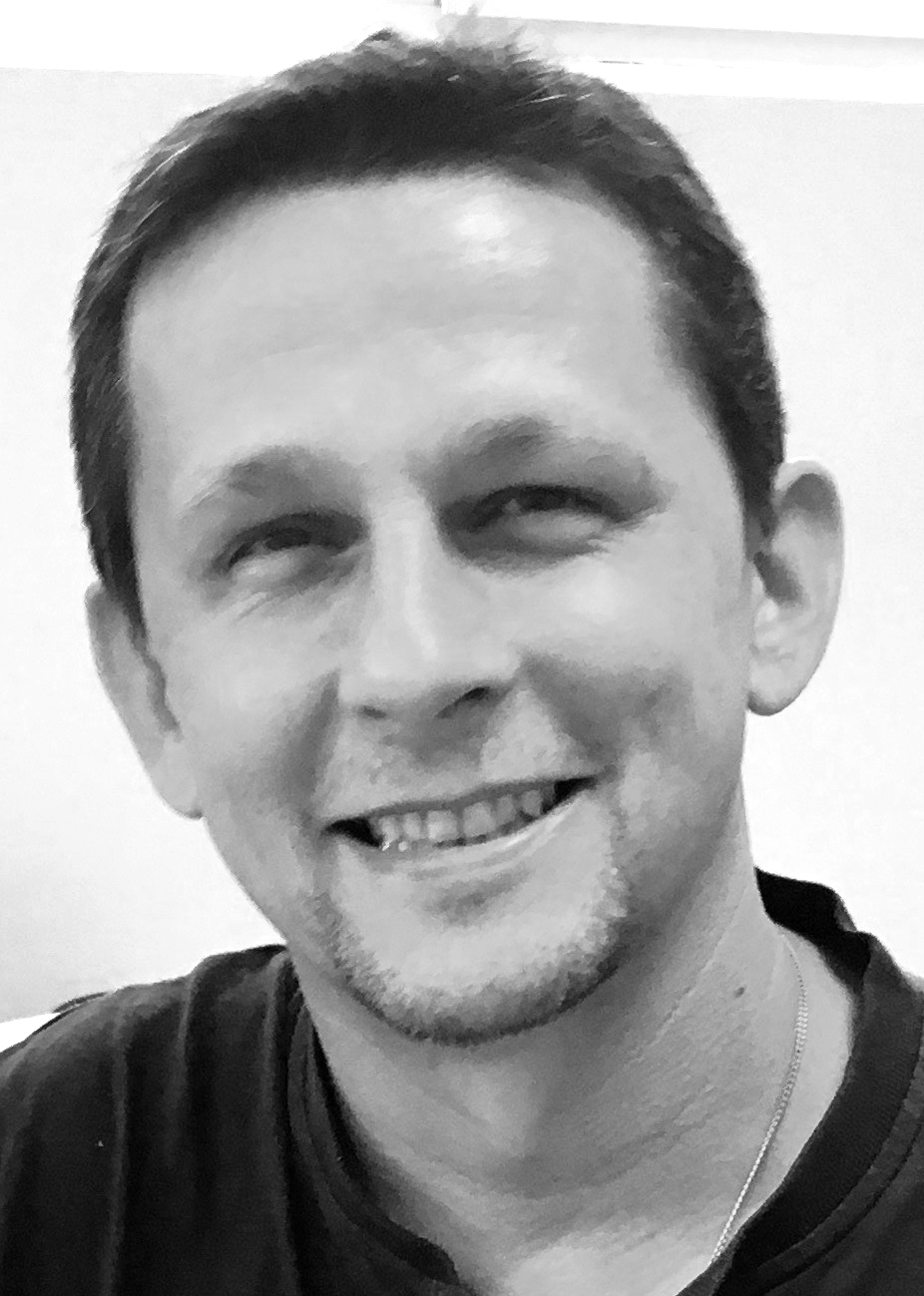}}]{Jean-Michel Portal} (M'87) is a Full Professor in  the Institute of Materials, Microelectronics and Nano-sciences of Provence, IM2NP at  Univerist\'e of Aix-Marseille. He received the Ph.D. degree in
1999 from University of Montpellier 2, France.
From 1999 to 2000, he was temporary researcher
at University of Montpellier 2 in the field of FPGA
design and test. From 2000 to 2008, he was assistant
professor at the Univ. of Provence, Polytech
Marseille, and conducted research activities in L2MP
in the field of Memory testing and diagnosis, test
structure design and design for manufacturing. In
this position he participated to industrial project
on non-volatile memory testing and diagnosis with ST Microelectronics. In 2008, he became Full Professor at Aix-Marseille
Univ. and since 2009 he heads the ``Memories Team'' of the IM2NP. His
research fields covers design for manufacturing and memory design, test and
reliability. 
\end{IEEEbiography}

\begin{IEEEbiography}[{\includegraphics[width=1in,height=1.25in,clip,keepaspectratio]{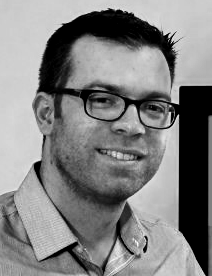}}]{Damien Querlioz} (M'08) is a CNRS Research Scientist at Univerist\'e Paris-Sud. He received his predoctoral education at Ecole Normale Sup\'erieure, Paris and his PhD from Universit\'e Paris-Sud in 2008. After postdoctoral appointments at Stanford University and CEA, he became a permanent researcher at the Centre for Nanoscience and Nanotechnology of Universit\'e Paris-Sud. He focuses on novel usages of emerging non-volatile memory, in particular relying on inspirations from biology and machine learning. Damien Querlioz coordinates the INTEGNANO interdisciplinary research group. In 2016, he was the recipient of an European Research Council Starting Grant to develop the concept of natively intelligent memory.   In 2017, he received the CNRS Bronze medal. 
\end{IEEEbiography}

\EOD

\end{document}